# Variations in Multi-Agent Actor-Critic Frameworks for Joint Optimizations in UAV Swarm Networks: Recent Evolution, Challenges, and Directions

Muhammad Morshed Alam
Dept. of Electrical and Electronic Engineering
AIUB
Dhaka, Bangladesh
dr.malam@aiub.edu

Muhammad Yeasir Aarafat
IT Research Institute
Chosun University
Gwangju 61452, South Korea
myarafat@chosun.ac.kr

Tamim Hossain
Dept. of Electrical and Electronic Engineering
AIUB
Dhaka, Bangladesh
tamim@aiub.edu

*Abstract*— Autonomous unmanned aerial vehicle (UAV) swarm networks (UAVSNs) can effectively execute surveillance, connectivity, and computing services to ground users (GUs). These missions require trajectory planning, UAV-GUs association, task offloading, next-hop selection, and resources such as transmit power, bandwidth, caching, and computing allocation to improve network performances. Owing to the highly dynamic topology, limited resources, and non-availability of global knowledge, optimizing network performance in UAVSNs is very intricate. Hence, it requires an adaptive joint optimization framework that can tackle both discrete and continuous decision variables to ensure optimal network performance under dynamic constraints. Multi-agent deep reinforcement learning-based adaptive actor-critic framework can efficiently address these problems. This paper investigates the recent evolutions of actor-critic frameworks to deal with joint optimization problems in UAVSNs. In addition, challenges and potential solutions are addressed as research directions.

## I. INTRODUCTION

Owing to the advantages of high maneuverability, autonomy, and flexibility, unmanned aerial vehicle (UAV) swarm networks (UAVSNs) can effectively execute various emergent missions. It includes simultaneous energy transfer and data collection from ground internet of things (IoT) devices [1]/ ground users (GUs), providing mobile edge computing services to resource-constrained IoT/GUs [2], aerial surveillance [3], and many more. These autonomous mission objectives are minimizing average age of information (AoI) [4], [5] or average delay in data collection and task execution [2], maximizing area coverage [6], maximizing fairness aware throughput [7], and minimizing energy consumption for both UAV and IoT/GUs [8]. Thus, achieving the above objectives in UAVSNs requires jointly optimizing several continuous and discrete decision variables. It may include controlling the continuous trajectory of UAVs [9], binary UAV-GUs association, determining binary or partial task-offloading, and continuous resource allocation (i.e. transmit power, bandwidth/timeslot, caching, and computing).

In these optimization processes, several mixed integer non-linear dynamic constraints are involved related to UAV's mobility, resource limitations (i.e., bandwidth, transmit power, onboard energy, and maximum queue size), UAV-GUs association, signal-to-interference-plus noise ratio (SINR) threshold, and minimum delay or AoI threshold. Furthermore, owing to the mobility of UAVs and GUs, network topology and channel state become very dynamic. Thus, the network state and decision space become very large. Additionally, the environment behavior becomes both cooperative and competitive in terms of collaborative trajectory planning and resource sharing in multiple layers along with multiple decision makers (i.e. UAVs, IoT devices/GUs, high altitude platforms, and satellites), as illustrated in Fig. 1.

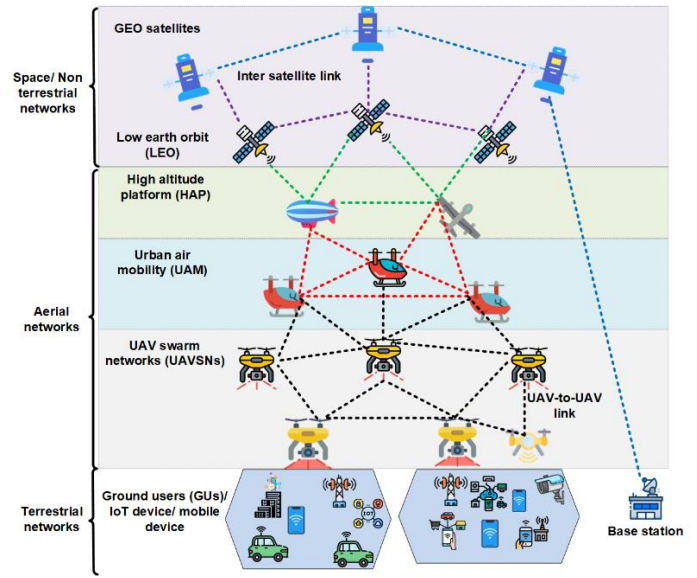

Fig. 1. An illustration of a UAVSN-assisted mission.

Thus, UAVSNs behavior can be treated as multi-agent systems [3], [10]. The action space with neighboring agents becomes highly coupled, more specifically, actions taken by the neighboring agents have a higher impact on the network utility of a particular agent. As a result, existing fully centralized model-based iterative algorithms are not suitable for this unknown and highly dynamic environment [2]. Here, collecting the centralized information on the dynamic topology is cumbersome and this outdated centralized information-based network optimization can trigger performance degradation. Furthermore, the computational complexity and control overhead increase with the increasing number of UAVs and GUs, which limits the scalability [2].

To combat the above-mentioned challenges, model-free data-driven multi-agent deep reinforcement learning (MA-DRL) can be efficiently utilized [11]. By leveraging the partially observable Markov decision process (POMDP), each agent can iteratively interact with the dynamic UAVSN to optimize the long-term cumulative reward based on their current observation. MA-DRL approaches can deal with multiple dynamic constraints by imposing penalty terms in the reward for choosing unfeasible actions by the agent without requiring convexity. Unlike the single-agent DRL algorithm, MA-DRL can avoid the environmental non-stationary problem by sharing observation-action via centralized training and distributed execution. Existing value-based MA-DRL approaches such as double deep Q-learning (DDQN) and QMIX [12] are suitable only for small-scale discrete action spaces. Hence, these algorithms are not suitable for large hybrid action spaces and extracting features from the dynamic UAVSNs. In contrast, policy gradient-based actor-critic learning can efficiently tackle large observation spaces and hybrid action spaces. Here, the actor-network is responsible for generating continuous and discreate actions. Consequently, critic network is responsible for evaluating the action by estimating the value function, which is further utilized to train the actor network and critic network [13]. The discrete action spaces such as binary task off-loading ratio, discrete frequency band selection, and binary UAV-GUs association can be mapped into continuous probability distribution using softmax activation function at the output layer of the actor-network. Then, during exploitation, the possible action can be sampled according to the maximum probability [14].

According to the existing literature, policy-based multi-agent actor-critic algorithms can be classified as off policy-based deep deterministic policy gradient (DDPG) [6], on policy-based trust region policy optimizations (TRPO) [15], and on policy-based proximal policy optimization (PPO) [4]. Among them, DDPG can handle only continuous action spaces such as trajectory control of UAVs or resource allocation in continuous domains. TRPO and PPO can tackle both continuous and discrete action spaces. Although these policy-based MA-DRL approaches have superior performance compared to the DDQN or QMIX, they encounter instability during policy updates due to the variability of the step-size parameter in policy optimization [4]. Small step sizes trigger slow learning, while large ones can degrade performance. To address this, TRPO defines a trusted region for updating policy using a complex second-order method. PPO simplifies it by employing a first-order method, which bounds policy updates within a range instead of TRPO's hard constraint. PPO has two key variants: PPO-penalty and PPO-clip. PPO-penalty replaces TRPO's hard constraint with a penalty in the objective function. In contrast, PPO-clip utilizes clipping mechanism to limit policy changes without a hard constraint [4].

*A. Motivation for This Paper*

Owing to the dynamic topology of UAVSNs, actor-critic-based MA-DRL algorithms are highly suitable for sequential decision-making. Nevertheless, an actor-critic network solely depending on a fully connected network (FCN) or multi-layer perceptron (MLP) cannot deal with time series sequential data such as UAV trajectories and channel states [2]. Additionally, FCN requires a higher number of learning parameters and this FCN-based learning is less scalable and not adaptive to environmental changes [16]. Fortunately, the recurrent neural network (RNN) such as long short-term memory (LSTM) and gated recurrent unit (GRU) can remember a finite amount of the most recent historical sequential data and utilize it to predict more accurate state in the next timeslot by mining the temporal relationship with the state information in the current timeslot [3]. Nevertheless, RNN-based LSTM or GRU-based encoder-decoder networks may encounter gradient vanishing problems during backpropagation and cannot efficiently deal with lengthy time-series data [3]. Moreover, RNN cannot adopt parallelization in feature extraction. The transformer-based encoder-decoder networks can overcome these limitations via multi-head attention mechanisms [17]. However, how to employ these networks in actor-critic frameworks to address joint optimization problems in dynamically resource-constrained UAVSNs, remains a research question. Although plenty of research has been conducted to address joint optimization problems in UAVSNs by leveraging various actor-critic-based MA-DRL frameworks, no in-depth survey has focused on the variations of actor-critic frameworks, their key limitations, research challenges, and potential solutions as future research directions. It motivates us to write this survey.

The remainder of this paper is arranged as follows. In Section II, the recent evolution of actor-critic frameworks in UAVSNs is briefly discussed in terms of the target optimization objective, constraints, POMDP formulation, modifications of the internal actor-critic network, policy gradient, loss function, and computational complexity. In Section III, research challenges and directions are summarized. Finally, this paper is concluded in Section IV.

II. RECENT EVOLUTION IN ACTOR-CRITIC FRAMEWORKS

In this Section, based on the internal neural network architecture, the recent evolution of actor-critic frameworks to deal with joint optimization problems in UAVSNs are classified as follows: Solely FCN or MLP-based, RNN-based, hybrid, and transformer-based encoder-decoder. Based on this classification each proposed framework is briefly reviewed in terms of their optimization objectives, constraints, POMDP formulation, training method, policy gradient, and computational complexity.

*A. Solely FCN or MLP-based*

This type of actor-critic framework utilizes only FCN or MLP in both actor and critic networks. It struggles to capture key features from dynamic UAVSNs, and training is not adaptive to varying numbers of UAVs or GUs. Thus, it has less scalability and higher computational complexity. In [18], minimizes the total system cost consisting of weighted energy consumption and delay-aware task priority-based utility function in GUs-UAVSNs assisted edge computing mission. They have considered binary task offloading decisions, continuous UAV mobility adjustment, transmit power, and computing resource allocation in the hybrid action space. To deal with hybrid action space, existing algorithms approximate either solely in discrete or continuous probability distributions, which may degrade optimization performance. Thus, considering the hybrid action space they have designed a learnable conditional variational auto-encoder (VAE), and decoder network based on FCN in the presence of Gaussian latent distribution in the encoder. This VAE can approximate better latent space for hybrid action representation in the continuous domain to ensure policy gradient for training the FCN-based actor-critic framework utilizing twin delayed MA-DDPG.

*B. RNN-based*

In RNN-based actor-critic frameworks, actor-critic networks are mainly constructed based on either LSTM or GRU cells. In some cases, LSTM or GRU-based state encoder-decoder is formed to predict the dynamic topology of UAVSNs. In [4], proposed a mean field game (MFG) hybrid PPO-based actor-critic framework to minimize average AoI by jointly optimizing UAV trajectories and data collection scheduling of GUs in hybrid action space. To predict the dynamic topology and stabilize the training process, this framework utilizes a learnable LSTM-based state encoder network. Then, LSTM-based state encoder output is fed into the two isolated actor networks to generate continuous and discrete actions related to UAV's trajectory and transmission scheduling of GUs, respectively. In [10], utilizes LSTM in both actor-critic networks to control the trajectory and maintain adaptive formation of UAVSNs in continuous action space (i.e., acceleration) in the presence of dense obstacles with the help of an improved potential field-based reward function. They have trained the model using twin delayed MA-DDPG. This model performs well owing to the advantages of capturing long-term dependencies in the sequential trajectory data by using LSTM.

*C. Hyrbid Structure-based*

Hybrid structure-based actor-critic frameworks are classified into four categories according to the type of neural network as follows:

*1) RNN-based Actor and Multi-head Attention-based Critic*

This type of actor-critic framework employs either LSTM or GRU in the actor-network to capture the sequential behavior of dynamic states. Subsequently, utilizes a multi-head cross attention-based critic network to precisely estimate the value-function. In [3], proposed a behavior-based mobility integrated LSTM-based actor and multi-head cross attention-based critic framework to maximize a multi-metric link utility to deal with packet routing problem in dynamic UAVSNs. They have considered the cross-layer design consisting of a large hybrid action space which includes collaborative trajectory control, frequency allocation, and next hop selection. The model is trained using distributed MA-DDPG consisting of two-hop neighbor information-based POMDP. The actor network is constructed based on LSTM-based three layers state encoding layer and an FCN. LSTM-based state encoding layers extracted key features from dynamic topology by detecting temporal continuity utilizing current and historical information. Subsequently, a multi-head cross attention-based critic network determines the precise value-function to train the actor network by paying attention to the neighboring agent's observation-action space according to their degree of influence. It provides training stability, overcomes environmental nonstationary, training adaptivity, and algorithm convergence.

In [2], a swarming behavior integrated GRU-based actor and a multi-head attention-based critic network is proposed for an air-ground integrated network. Here, mobile edge computing-enabled UAVSN acts as agents to jointly minimize the task execution delay for mobile GUs and energy consumption for UAVs by controlling UAV's trajectory, UAV-GUs association, partial task off-loading ratio, computing, and bandwidth allocation for G2U links under several key constraints. They introduced behavior-based motion model and adaptive penalty terms to enhance decision making. The GRU-based actor network is capable of coping with dynamic observation of each agent. Similarly, the multi-head attention-based critic network utilizes cross-attention mechanism to identify the influence of other neighboring agent's dynamic observation-action space for each individual UAV-agents. Moreover, it provides training adaptivity to different number of UAVs and GUs, system scalability, and helps to avoid environmental nonstationary.

*2) Multi-head Attention-based*

This type of actor-critic frameworks, employed multi-head attention mechansims in both actor-critic networks to provide both self and cross attention in multi-agent environments. In [7], investigates a joint optimization on trajectory planning and communication design in an air-ground communication systems by adopting actor-critic-based heterogenous collaborative decision making algorithm considering UAV base stations and GUs as intelligent agent. Thus, the action space for UAV base stations includes trajectory planning and bandwidth allocation for G2U links to maximize long-term fair throughput. Similarly, the action space for GUs includes binary UAV-GUs association to compete for limited UAV resources in each timeslot for maximizing their own long-term throughput. To quickly identify the correlation with other agent's complex observation-action space, enhance the model adaptability and the decision-making process, they introduced a learnable multi-head attention mechanism in both actor and critic network of all agents.

*3) Convolution Neural Network (CNN)-based*

In joint optimization problems typically the observation and action space are very large and dynamic. Thus, convolution neural networks (CNN) can be utilized to extract the key features from the observation-action spaces by reducing the dimensionality. Then, CNN followed by LSTM or GRU layers and MLP can learn temporal dependencies more efficiently via utilizing historical knowledge of sequential data. Nevertheless, the computational complexity and vanishing gradient problem are key limitations of this network. In [5], utilizes federated heterogeneous actor-critic frameworks to minimize AoI for computing tasks in a multi-agent mobile GUs-UAV-cloud environment by jointly considering computation offloading, UAV trajectory, and bandwidth allocation. Since tasks arrived at mobile GUs following Poisson distribution, dynamic task queue, and topology, they have considered CNN and MLP-based actor networks. CNN can extract the areas of GUs from the dynamic observation that has higher computation demand and AoI, thus, accordingly the trajectory of UAVs can be designed.

*4) Graph Attention Network (GAT)-based*

In this type of framework, actor-critic network is designed based on a graph attention network (GAT). GUs and UAVSNs topology can be represented as a dynamic graph attention network and it can be efficiently trained using the MA-DRL method [19]. The learnable attention weights in GAT facilitate UAVs to pay more attention to the nearby UAVs according to their degree of influence in decision-making via communication range restricted partial observation [20]. In [21], addressed the joint non-divisible task offloading and resource allocation (i.e., bandwidth and computing) in a hierarchical model considering mobile devices, edge, and cloud servers. This model leverages a learnable adaptive attention-weighted graph RNN-based state encoder utilizing the local knowledge followed by an actor-critic framework. Owing to the learnable state encoding using spatio-temporal graph RNN, the model can generalize different numbers of mobile devices and edge servers via fixed-size embedding, which leads to higher scalability and training adaptability to the unknown workload on the edge server.

*D. Transformer-based Encoder-Decoder*

This type of framework utilizes a transformer-based encoder-decoder structure in both actor and critic networks to learn from long observation-action spaces in UAVSNs. It can avoid vanishing gradient problems, remember long sequences, and train adaptivity to different sizes of networks via fixed-size embedding. However, it has a higher computational complexity. In [14], proposed transformer-based encoder-decoder actor-critic frameworks to jointly minimize the long-term average weighted sum of task execution-delay and energy consumption for multi-UAV and multi-edge server networks. Here the action space consists of a binary task offloading policy and continuous resource allocation that includes transmit power, bandwidth, and computing resources. The model employed a DDPG-based policy gradient during training and adopted a transformer-based encoder-decoder structure to accelerate algorithm convergence due to the capabilities of handling large state-action decision space and avoidance of vanishing gradient problems. Furthermore, owing to the encoder-decoder-based key feature extractions, it is adaptive to different sizes of UAVs and task attributes.

In [6], investigates an area coverage problem for multi-UAVs by utilizing transformer-based state encoder-based actor-critic frameworks. The model can make long-term trajectory decisions for UAVs to maximize fairness-aware area coverage toward points of interest. The transformer-based encoder network considers the observation of the neighbor UAV's state as key and value, each UAV's state as the query to generate a state embedding feature vector. This extracted feature vector from the dynamic environment is fed into an MLP-based actor-critic framework for training the model. Through feature extraction, the model is adaptive to different state features and different numbers of UAVs. Additionally, instead of random initialization of learnable weights in actor-critic networks, the designed model adopted a baseline-assisted pre-training mechanism which can accelerate algorithm convergence. The above discussions in this Section are summarized in Table I.

TABLE I. SUMMARY OF ACTOR-CRITIC FRAMEWORKS TO DEAL WITH JOINT OPTIMIZATION PROBLEMS IN UAVSNs.

| Type | Ref. | Optimization objective | Action space | Key contributions | Advantages | Limitations |
|---|---|---|---|---|---|---|
| Solely FCN or MLP-based | [18] | Minimizes total cost consisting of energy consumption and delay-aware task priority-based utility in UAVSNs assisted edge computing mission for GUs | Binary task offloading, transmit power of GUs, UAV position and computing resource | Designing a conditional VAE-decoder-based continuous hybrid latent space | VAE provides better approximation for hybrid action space in continuous domain and ensure policy gradient for twin delayed MA-DDPG | FCN has higher computational complexity, less scalability, and less adaptive to dynamic topology |
| RNN-based | [4] | Minimizes average AoI in UAVSNs assisted data collection mission from GUs | Scheduling data transmission for GUs and UAVs mobility | Solving a MFG by applying PPO-based continuous-discrete actor-critic | To predict the dynamic topology and stabilize the training process, utilizes a learnable LSTM-based state encoder network | LSTM faces vanishing gradient problem, and defined MFG can support only homogenous environment |
| | [10] | Trajectory planning for UAVSNs while maintaining formation in an unknown dense obstacle environment | Acceleration of UAVs | Utilizes LSTM in both actor-critic networks adaptively control the trajectory | Improved potential field-based reward helps quickly achieve optimal policy for twin delayed MA-DDPG | LSTM cannot efficiently remember long trajectory sequences |
| Hybrid structure-based | [3] | Maximizes a multi-metric link utility to route data packet in UAVSNs by utilizing two hop neighbor information | UAVs trajectory control, frequency band, and next hop selection | A swarming behavior-based mobility integrated LSTM-based actor and multi-head cross attention-based critic framework is proposed | State decoupling in LSTM-based actor network helps to achieve better policy, and avoid environmental non-stationarity via multi-head cross attention | LSTM-based actor may face overfitting and vanishing gradient problem |
| | [2] | Minimizes energy consumption of UAVs and delay in air-ground integrated edge computing networks | Trajectory of UAVs, UAV-GUs association, bandwidth, and computing resource allocation | A behavior integrated GRU-based actor and a multi-head attention-based critic network is proposed | GRU-based actor network is capable of coping with dynamic observation. Multi-head attention-based critic network utilizes cross-attention to identify influence of other neighboring agents. | Vanishing gradient problem and struggle to remember large observation |
| | [7] | Maximizes throughput for G2U links to provide communication service | UAVs: trajectory planning and bandwidth allocation, and GUs: binary UAV-GUs association | Proposed a heterogenous collaborative actor-critic framework based on multi-head attention-based | Learnable muti-head attention enhances the model's adaptability and the decision-making process | Higher computational complexity |
| | [5] | Minimizes AoI for computing stochastic tasks in a multi-agent mobile GUs-UAV-cloud environment | Computation offloading, UAV trajectory, and bandwidth allocation | Proposed a federated heterogeneous CNN-based actor-critic frameworks | CNN can extract the areas of GUs that have higher computation demand and AoI. | Higher computational complexity |
| | [21] | Minimizes long-term system costs consisting of energy and delay in a mobile device-edge server-cloud networks | Binary task offloading, computing, and bandwidth allocation | Leverages attention-weighted graph RNN-based state encoder utilizing the local knowledge | Highly scalable, utilizes log probability distribution-based policy gradient to support hybrid action space and fully decentralized | Higher computational complexity |
| Transformer-based Encoder-Decoder | [14] | Minimizes long-term average weighted task execution delay and energy consumption for multi-UAV and multi-edge server networks | Binary task offloading, transmit power, bandwidth, and computing resources | Proposed transformer-based encoder-decoder actor-critic frameworks based on MA-DDPG | Can remember long observation-action sequences via attention mechanism, highly scalable, and higher training adaptivity for dynamic UAVSNs | Higher computational complexity, critic network provides only self-attention and a long training time |

## III. RESEARCH CHALLENGES AND DIRECTIONS

Based on the discussion in Section II, research challenges and solutions are summarized as follows:

### A. Physics Informed Neural Networks for Trajectory Control of UAVs

Existing model-free data-driven algorithms applied to control the trajectory of UAVs are trained neural networks based on either supervised learning or reinforcement learning. Physics-informed neural networks (PINNs) introduce UAVs motion dynamics in the form of partial differential equations or motion models [3] into the neural networks as loss function, augmenting the accuracy, generalization, and robustness of the model by ensuring predictions adhere to known physical principles. In addition, PINNs can assist in tackling uncertainties related to wind disturbance, UAV failures, or unexpected obstacles by embedding robust physical laws. Hence, PINNs can provide better training and algorithm convergence with fewer data samples considering real-life scenarios, which need further exploration.

### B. Model Traning and Heterogeneous Environment

Existing multi-agent actor-critic frameworks are adopted centralized training and decentralized or distributed execution. However, considering the limited communication range of GUs and UAVs distributed or federated training still requires further study [3]. An intelligent agent can use an attention mechanism to obtain the key features in multi-agent scenarios via a cross-attention mechanism in the form of query, key, and value by taking only partial observation-action as input. The hierarchical space air-ground integrated networks introduce multiple links and decision-makers as intelligent agents at each layer. Hence, a heterogenous collaborative-competitive environment appears which requires special neural networks for each agent according to its resource capacity to make the algorithm more robust and practical. The environment should be high fidelity environment for training, which can be achieved by employing digital twin-assisted training [8] and introducing noise not only in action space but also in state space to introduce uncertainties.

### C. Transformer-based Encoder-Decoder Multi-agent Actor-Critic Frameworks

Actor networks require self-attention on its long observation sequences to predict the dynamic topology state by generating fixed-size state embedding and mapping it into the optimal action. In contrast, critic networks require cross attention to pay attention to not only its own observation-action spaces but also to the neighboring agents in a multi-agent cooperative-competitive environment to precisely estimate the value function [2]. This type of framework can deal with joint trajectory control, task offloading, and resource allocation problems that require hybrid action spaces in UAVSNs. Additionally, in space-air-ground integrated networks, computationally intensive and delay critical tasks of GUs are offloaded based on binary or partially only to the hierarchical locations (UAVs, high altitude platforms, or low earth orbit satellites). Here, neighboring UAVs might have ideal resources that can be traced by paying attention to their observations to introduce both vertical and horizontal offloading. Thus, low complexity transformer-based self-attention encoder-decoder in actor and cross-attention-based encoder-decoder in critic networks require further study to deal with these problems.

*D. Channel State Estimation in UAVSNs*

Owing to mobility and dense obstacles both UAV-to-UAV and UAV-to-GU links are highly dynamic and frequent link outages may occur. To adopt the mm-wave and free space optic in UAVSNs, the power allocation, beamforming, beam training, and controlling the phase shift from the intelligent reflecting surface, require accurate channel state information by sending pilot signals. Nevertheless, the frequent transmission of pilot signals triggers a waste of resources and delays in communication. Thus, channel state information in highly dynamic UAVSNs can be predicted by designing transformer-based encoder-decoder frameworks with less pilot transmission, which can be trained further by leveraging reinforcement learning.

IV. CONCLUSION

In this paper, the evolution of existing state-of-the-art multi-agent actor-critic frameworks is reviewed based on their internal neural network structure that is employed to solve various joint optimization problems in UAVSNs along with their advantages and limitations. Additionally, open issues and potential solutions are also addressed. We believe this article will assist researchers in enhancing network performance in UAVSNs by employing data-driven MA-DRL-based actor-critic frameworks.

ACKNOWLEDGMENT

The authors thank the editor and anonymous referees for their comments that helped improve the quality of this manuscript.